\pdfoutput=1
\documentclass[preprint,prd,tightenlines, superscriptaddress]{revtex4}

\usepackage{graphicx} 
\usepackage{dcolumn}  
\usepackage{colordvi}
\usepackage{color}
\usepackage{epstopdf}
\usepackage{subfig}
\usepackage{caption}
\usepackage{amssymb}
\usepackage{url}
\graphicspath{{ps}}
\usepackage{hyperref}
\usepackage{tabularx}

\renewcommand{\arraystretch}{1.1}

\input belle2sym.tex

\def\X           {\ensuremath{X}\xspace}

\def \Y       {\ensuremath{\Upsilon(4S)}\xspace}

\begin{document}


\def\belletwo {{Belle II}\xspace}
\def\itbelletwo {{\it {Belle II}}\xspace}
\def\phaseiii {{Phase III}\xspace}
\def\itphaseiii {{\it {Phase III}}\xspace}

\newcommand\logten{\ensuremath{\log_{10}\;}}


\def\lint {34.6 \invfb}
\def\procversion {{\it {proc11 + prompt}}\xspace}
\def\mcversion {{MC13a}\xspace}
\def\release {{\it {light-2002-ichep}}\xspace}
\def\feitraining {{\it {FEIv4\_2020\_MC13\_release\_04\_01\_01}}\xspace}
\def\sigprobcut {{$0.001$}\xspace}


\vspace*{-3\baselineskip}
\resizebox{!}{3cm}{\includegraphics{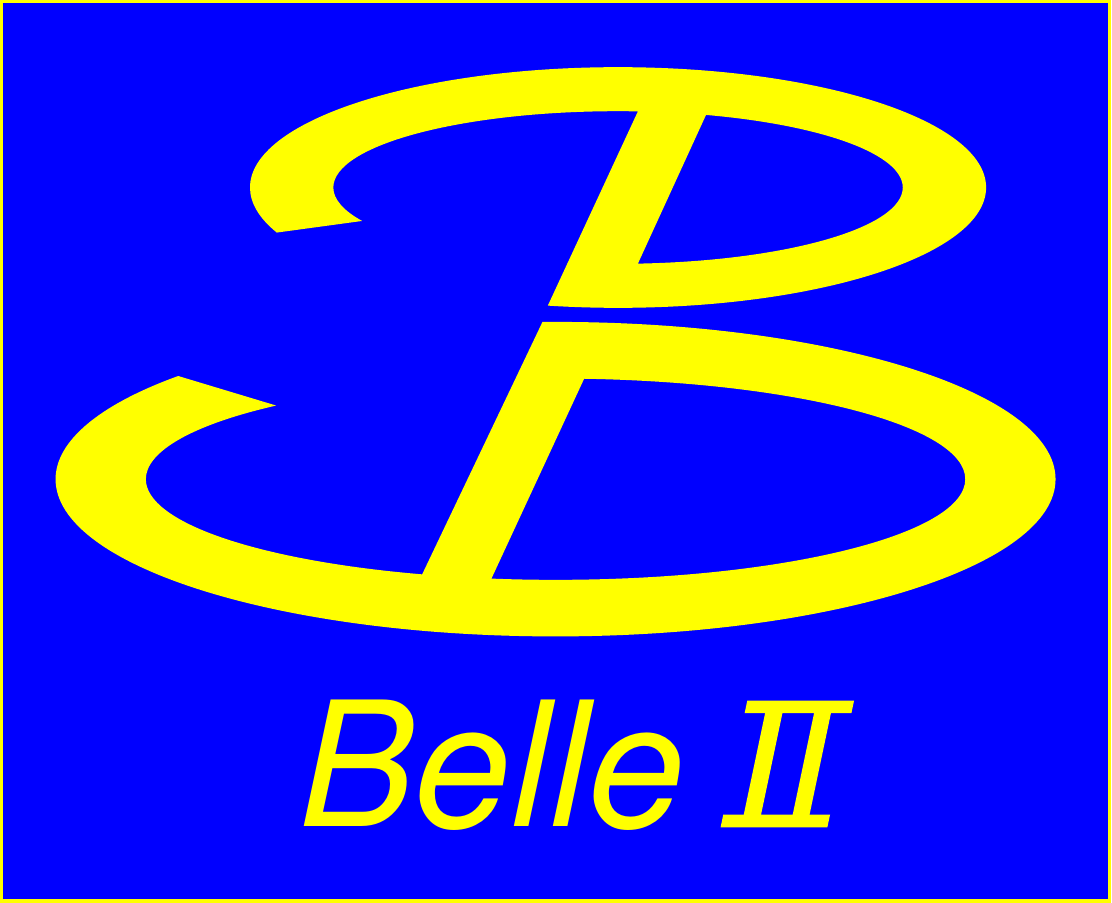}}

\vspace*{-5\baselineskip}
\begin{flushright}
\vspace{1em}
BELLE2-CONF-PH-2020-007\\
\today

\end{flushright}

\vspace{2em}
\title {\LARGE Exclusive $B^0 \to \pi^- \ell^+ \nu_\ell$ Decays with Hadronic Full Event Interpretation Tagging in 34.6\invfb of Belle II Data\\
\vspace{1em}
}



\newcommand{\instSinica}{Academia Sinica, Taipei 11529, Taiwan}
\newcommand{\instCPPM}{Aix Marseille Universit\'{e}, CNRS/IN2P3, CPPM, 13288 Marseille, France}
\newcommand{\instBeihang}{Beihang University, Beijing 100191, China}
\newcommand{\instBUAP}{Benemerita Universidad Autonoma de Puebla, Puebla 72570, Mexico}
\newcommand{\instBNL}{Brookhaven National Laboratory, Upton, New York 11973, U.S.A.}
\newcommand{\instBINP}{Budker Institute of Nuclear Physics SB RAS, Novosibirsk 630090, Russian Federation}
\newcommand{\instCMU}{Carnegie Mellon University, Pittsburgh, Pennsylvania 15213, U.S.A.}
\newcommand{\instCinvestavIPN}{Centro de Investigacion y de Estudios Avanzados del Instituto Politecnico Nacional, Mexico City 07360, Mexico}
\newcommand{\instPrague}{Faculty of Mathematics and Physics, Charles University, 121 16 Prague, Czech Republic}
\newcommand{\instChiangMai}{Chiang Mai University, Chiang Mai 50202, Thailand}
\newcommand{\instChiba}{Chiba University, Chiba 263-8522, Japan}
\newcommand{\instChonnam}{Chonnam National University, Gwangju 61186, South Korea}
\newcommand{\instConacyt}{Consejo Nacional de Ciencia y Tecnolog\'{\i}a, Mexico City 03940, Mexico}
\newcommand{\instDESY}{Deutsches Elektronen--Synchrotron, 22607 Hamburg, Germany}
\newcommand{\instDuke}{Duke University, Durham, North Carolina 27708, U.S.A.}
\newcommand{\instITAR}{Institute of Theoretical and Applied Research (ITAR), Duy Tan University, Hanoi 100000, Vietnam}
\newcommand{\instENEA}{ENEA Casaccia, I-00123 Roma, Italy}
\newcommand{\instEri}{Earthquake Research Institute, University of Tokyo, Tokyo 113-0032, Japan}
\newcommand{\instJuelich}{Forschungszentrum J\"{u}lich, 52425 J\"{u}lich, Germany}
\newcommand{\instFuJen}{Department of Physics, Fu Jen Catholic University, Taipei 24205, Taiwan}
\newcommand{\instFudan}{Key Laboratory of Nuclear Physics and Ion-beam Application (MOE) and Institute of Modern Physics, Fudan University, Shanghai 200443, China}
\newcommand{\instGoettingen}{II. Physikalisches Institut, Georg-August-Universit\"{a}t G\"{o}ttingen, 37073 G\"{o}ttingen, Germany}
\newcommand{\instGifu}{Gifu University, Gifu 501-1193, Japan}
\newcommand{\instSOKENDAI}{The Graduate University for Advanced Studies (SOKENDAI), Hayama 240-0193, Japan}
\newcommand{\instGyeongsang}{Gyeongsang National University, Jinju 52828, South Korea}
\newcommand{\instHanyang}{Department of Physics and Institute of Natural Sciences, Hanyang University, Seoul 04763, South Korea}
\newcommand{\instKEK}{High Energy Accelerator Research Organization (KEK), Tsukuba 305-0801, Japan}
\newcommand{\instJPARC}{J-PARC Branch, KEK Theory Center, High Energy Accelerator Research Organization (KEK), Tsukuba 305-0801, Japan}
\newcommand{\instHSE}{Higher School of Economics (HSE), Moscow 101000, Russian Federation}
\newcommand{\instIISER}{Indian Institute of Science Education and Research Mohali, SAS Nagar, 140306, India}
\newcommand{\instIITBhubaneswar}{Indian Institute of Technology Bhubaneswar, Satya Nagar 751007, India}
\newcommand{\instIITGuwahati}{Indian Institute of Technology Guwahati, Assam 781039, India}
\newcommand{\instIITHyderabad}{Indian Institute of Technology Hyderabad, Telangana 502285, India}
\newcommand{\instIITMadras}{Indian Institute of Technology Madras, Chennai 600036, India}
\newcommand{\instIndiana}{Indiana University, Bloomington, Indiana 47408, U.S.A.}
\newcommand{\instIHEPRussia}{Institute for High Energy Physics, Protvino 142281, Russian Federation}
\newcommand{\instHEPHYVienna}{Institute of High Energy Physics, Vienna 1050, Austria}
\newcommand{\instIHEPChina}{Institute of High Energy Physics, Chinese Academy of Sciences, Beijing 100049, China}
\newcommand{\instChennai}{Institute of Mathematical Sciences, Chennai 600113, India}
\newcommand{\instIPP}{Institute of Particle Physics (Canada), Victoria, British Columbia V8W 2Y2, Canada}
\newcommand{\instIOP}{Institute of Physics, Vietnam Academy of Science and Technology (VAST), Hanoi, Vietnam}
\newcommand{\instIFIC}{Instituto de Fisica Corpuscular, Paterna 46980, Spain}
\newcommand{\instFrascati}{INFN Laboratori Nazionali di Frascati, I-00044 Frascati, Italy}
\newcommand{\instNapoliINFN}{INFN Sezione di Napoli, I-80126 Napoli, Italy}
\newcommand{\instPadovaINFN}{INFN Sezione di Padova, I-35131 Padova, Italy}
\newcommand{\instPerugiaINFN}{INFN Sezione di Perugia, I-06123 Perugia, Italy}
\newcommand{\instPisaINFN}{INFN Sezione di Pisa, I-56127 Pisa, Italy}
\newcommand{\instRomaINFN}{INFN Sezione di Roma, I-00185 Roma, Italy}
\newcommand{\instRomaTreINFN}{INFN Sezione di Roma Tre, I-00146 Roma, Italy}
\newcommand{\instTorinoINFN}{INFN Sezione di Torino, I-10125 Torino, Italy}
\newcommand{\instTriesteINFN}{INFN Sezione di Trieste, I-34127 Trieste, Italy}
\newcommand{\instJAEA}{Advanced Science Research Center, Japan Atomic Energy Agency, Naka 319-1195, Japan}
\newcommand{\instMainz}{Johannes Gutenberg-Universit\"{a}t Mainz, Institut f\"{u}r Kernphysik, D-55099 Mainz, Germany}
\newcommand{\instGiessen}{Justus-Liebig-Universit\"{a}t Gie\ss{}en, 35392 Gie\ss{}en, Germany}
\newcommand{\instKarlsruhe}{Institut f\"{u}r Experimentelle Teilchenphysik, Karlsruher Institut f\"{u}r Technologie, 76131 Karlsruhe, Germany}
\newcommand{\instKennesaw}{Kennesaw State University, Kennesaw, Georgia 30144, U.S.A.}
\newcommand{\instKitasato}{Kitasato University, Sagamihara 252-0373, Japan}
\newcommand{\instKISTI}{Korea Institute of Science and Technology Information, Daejeon 34141, South Korea}
\newcommand{\instKorea}{Korea University, Seoul 02841, South Korea}
\newcommand{\instKSU}{Kyoto Sangyo University, Kyoto 603-8555, Japan}
\newcommand{\instKyotoU}{Kyoto University, Kyoto 606-8501, Japan}
\newcommand{\instKyungpook}{Kyungpook National University, Daegu 41566, South Korea}
\newcommand{\instLPI}{P.N. Lebedev Physical Institute of the Russian Academy of Sciences, Moscow 119991, Russian Federation}
\newcommand{\instLNNU}{Liaoning Normal University, Dalian 116029, China}
\newcommand{\instLMU}{Ludwig Maximilians University, 80539 Munich, Germany}
\newcommand{\instLuther}{Luther College, Decorah, Iowa 52101, U.S.A.}
\newcommand{\instMNITJaipur}{Malaviya National Institute of Technology Jaipur, Jaipur 302017, India}
\newcommand{\instMPP}{Max-Planck-Institut f\"{u}r Physik, 80805 M\"{u}nchen, Germany}
\newcommand{\instMPGHLL}{Semiconductor Laboratory of the Max Planck Society, 81739 M\"{u}nchen, Germany}
\newcommand{\instMcGill}{McGill University, Montr\'{e}al, Qu\'{e}bec, H3A 2T8, Canada}
\newcommand{\instMETU}{Middle East Technical University, 06531 Ankara, Turkey}
\newcommand{\instMEPhI}{Moscow Physical Engineering Institute, Moscow 115409, Russian Federation}
\newcommand{\instNagoya}{Graduate School of Science, Nagoya University, Nagoya 464-8602, Japan}
\newcommand{\instNagoyaKMI}{Kobayashi-Maskawa Institute, Nagoya University, Nagoya 464-8602, Japan}
\newcommand{\instNagoyaIAR}{Institute for Advanced Research, Nagoya University, Nagoya 464-8602, Japan}
\newcommand{\instNaraWu}{Nara Women's University, Nara 630-8506, Japan}
\newcommand{\instUNAM}{National Autonomous University of Mexico, Mexico City, Mexico}
\newcommand{\instNTUTaiwan}{Department of Physics, National Taiwan University, Taipei 10617, Taiwan}
\newcommand{\instNUUTaiwan}{National United University, Miao Li 36003, Taiwan}
\newcommand{\instKrakow}{H. Niewodniczanski Institute of Nuclear Physics, Krakow 31-342, Poland}
\newcommand{\instNiigata}{Niigata University, Niigata 950-2181, Japan}
\newcommand{\instNSU}{Novosibirsk State University, Novosibirsk 630090, Russian Federation}
\newcommand{\instOkinawa}{Okinawa Institute of Science and Technology, Okinawa 904-0495, Japan}
\newcommand{\instOsakaCity}{Osaka City University, Osaka 558-8585, Japan}
\newcommand{\instRCNP}{Research Center for Nuclear Physics, Osaka University, Osaka 567-0047, Japan}
\newcommand{\instPNNL}{Pacific Northwest National Laboratory, Richland, Washington 99352, U.S.A.}
\newcommand{\instPanjab}{Panjab University, Chandigarh 160014, India}
\newcommand{\instPeking}{Peking University, Beijing 100871, China}
\newcommand{\instPanjabPAU}{Punjab Agricultural University, Ludhiana 141004, India}
\newcommand{\instRIKENMSL}{Meson Science Laboratory, Cluster for Pioneering Research, RIKEN, Saitama 351-0198, Japan}
\newcommand{\instRIKEN}{Theoretical Research Division, Nishina Center, RIKEN, Saitama 351-0198, Japan}
\newcommand{\instXavier}{St. Francis Xavier University, Antigonish, Nova Scotia, B2G 2W5, Canada}
\newcommand{\instSeoul}{Seoul National University, Seoul 08826, South Korea}
\newcommand{\instShandong}{Shandong University, Jinan 250100, China}
\newcommand{\instSPU}{Showa Pharmaceutical University, Tokyo 194-8543, Japan}
\newcommand{\instSoochow}{Soochow University, Suzhou 215006, China}
\newcommand{\instSoongsil}{Soongsil University, Seoul 06978, South Korea}
\newcommand{\instLjubljanaJSI}{J. Stefan Institute, 1000 Ljubljana, Slovenia}
\newcommand{\instKyiv}{Taras Shevchenko National Univ. of Kiev, Kiev, Ukraine}
\newcommand{\instTata}{Tata Institute of Fundamental Research, Mumbai 400005, India}
\newcommand{\instTUM}{Department of Physics, Technische Universit\"{a}t M\"{u}nchen, 85748 Garching, Germany}
\newcommand{\instECUTUM}{Excellence Cluster Universe, Technische Universit\"{a}t M\"{u}nchen, 85748 Garching, Germany}
\newcommand{\instTelAviv}{Tel Aviv University, School of Physics and Astronomy, Tel Aviv, 69978, Israel}
\newcommand{\instToho}{Toho University, Funabashi 274-8510, Japan}
\newcommand{\instTohoku}{Department of Physics, Tohoku University, Sendai 980-8578, Japan}
\newcommand{\instTitech}{Tokyo Institute of Technology, Tokyo 152-8550, Japan}
\newcommand{\instTokyoMetropolitan}{Tokyo Metropolitan University, Tokyo 192-0397, Japan}
\newcommand{\instUAS}{Universidad Autonoma de Sinaloa, Sinaloa 80000, Mexico}
\newcommand{\instNapoliUNIV}{Dipartimento di Scienze Fisiche, Universit\`{a} di Napoli Federico II, I-80126 Napoli, Italy}
\newcommand{\instNapoliUNIVA}{Dipartimento di Agraria, Universit\`{a} di Napoli Federico II, I-80055 Portici (NA), Italy}
\newcommand{\instPadovaUNIV}{Dipartimento di Fisica e Astronomia, Universit\`{a} di Padova, I-35131 Padova, Italy}
\newcommand{\instPerugiaUNIV}{Dipartimento di Fisica, Universit\`{a} di Perugia, I-06123 Perugia, Italy}
\newcommand{\instPisaUNIV}{Dipartimento di Fisica, Universit\`{a} di Pisa, I-56127 Pisa, Italy}
\newcommand{\instRomaUNIV}{Universit\`{a} di Roma ``La Sapienza,'' I-00185 Roma, Italy}
\newcommand{\instRomaTreUNIV}{Dipartimento di Matematica e Fisica, Universit\`{a} di Roma Tre, I-00146 Roma, Italy}
\newcommand{\instTorinoUNIV}{Dipartimento di Fisica, Universit\`{a} di Torino, I-10125 Torino, Italy}
\newcommand{\instTriesteUNIV}{Dipartimento di Fisica, Universit\`{a} di Trieste, I-34127 Trieste, Italy}
\newcommand{\instMontreal}{Universit\'{e} de Montr\'{e}al, Physique des Particules, Montr\'{e}al, Qu\'{e}bec, H3C 3J7, Canada}
\newcommand{\instIJCLab}{Universit\'{e} Paris-Saclay, CNRS/IN2P3, IJCLab, 91405 Orsay, France}
\newcommand{\instIPHC}{Universit\'{e} de Strasbourg, CNRS, IPHC, UMR 7178, 67037 Strasbourg, France}
\newcommand{\instAdelaide}{Department of Physics, University of Adelaide, Adelaide, South Australia 5005, Australia}
\newcommand{\instBonn}{University of Bonn, 53115 Bonn, Germany}
\newcommand{\instUBC}{University of British Columbia, Vancouver, British Columbia, V6T 1Z1, Canada}
\newcommand{\instCincinnati}{University of Cincinnati, Cincinnati, Ohio 45221, U.S.A.}
\newcommand{\instFlorida}{University of Florida, Gainesville, Florida 32611, U.S.A.}
\newcommand{\instHamburg}{University of Hamburg, 20148 Hamburg, Germany}
\newcommand{\instHawaii}{University of Hawaii, Honolulu, Hawaii 96822, U.S.A.}
\newcommand{\instHeidelberg}{University of Heidelberg, 68131 Mannheim, Germany}
\newcommand{\instLjubljanaUniLJ}{Faculty of Mathematics and Physics, University of Ljubljana, 1000 Ljubljana, Slovenia}
\newcommand{\instLouisville}{University of Louisville, Louisville, Kentucky 40292, U.S.A.}
\newcommand{\instMalaya}{National Centre for Particle Physics, University Malaya, 50603 Kuala Lumpur, Malaysia}
\newcommand{\instLjubljanaUM}{University of Maribor, 2000 Maribor, Slovenia}
\newcommand{\instMelbourne}{School of Physics, University of Melbourne, Victoria 3010, Australia}
\newcommand{\instMississippi}{University of Mississippi, University, Mississippi 38677, U.S.A.}
\newcommand{\instUOM}{University of Miyazaki, Miyazaki 889-2192, Japan}
\newcommand{\instNovaGorica}{University of Nova Gorica, 5000 Nova Gorica, Slovenia}
\newcommand{\instPittsburgh}{University of Pittsburgh, Pittsburgh, Pennsylvania 15260, U.S.A.}
\newcommand{\instUSTC}{University of Science and Technology of China, Hefei 230026, China}
\newcommand{\instSAlabama}{University of South Alabama, Mobile, Alabama 36688, U.S.A.}
\newcommand{\instSCarolina}{University of South Carolina, Columbia, South Carolina 29208, U.S.A.}
\newcommand{\instSydney}{School of Physics, University of Sydney, New South Wales 2006, Australia}
\newcommand{\instTabuk}{Department of Physics, Faculty of Science, University of Tabuk, Tabuk 71451, Saudi Arabia}
\newcommand{\instUTokyo}{Department of Physics, University of Tokyo, Tokyo 113-0033, Japan}
\newcommand{\instIPMU}{Kavli Institute for the Physics and Mathematics of the Universe (WPI), University of Tokyo, Kashiwa 277-8583, Japan}
\newcommand{\instVictoria}{University of Victoria, Victoria, British Columbia, V8W 3P6, Canada}
\newcommand{\instVPI}{Virginia Polytechnic Institute and State University, Blacksburg, Virginia 24061, U.S.A.}
\newcommand{\instWayneState}{Wayne State University, Detroit, Michigan 48202, U.S.A.}
\newcommand{\instYamagata}{Yamagata University, Yamagata 990-8560, Japan}
\newcommand{\instYerevan}{Alikhanyan National Science Laboratory, Yerevan 0036, Armenia}
\newcommand{\instYonsei}{Yonsei University, Seoul 03722, South Korea}
\affiliation{\instCPPM}
\affiliation{\instBeihang}
\affiliation{\instBNL}
\affiliation{\instBINP}
\affiliation{\instCMU}
\affiliation{\instCinvestavIPN}
\affiliation{\instPrague}
\affiliation{\instChiangMai}
\affiliation{\instChiba}
\affiliation{\instChonnam}
\affiliation{\instConacyt}
\affiliation{\instDESY}
\affiliation{\instDuke}
\affiliation{\instITAR}
\affiliation{\instEri}
\affiliation{\instJuelich}
\affiliation{\instFuJen}
\affiliation{\instFudan}
\affiliation{\instGoettingen}
\affiliation{\instGifu}
\affiliation{\instSOKENDAI}
\affiliation{\instGyeongsang}
\affiliation{\instHanyang}
\affiliation{\instKEK}
\affiliation{\instJPARC}
\affiliation{\instHSE}
\affiliation{\instIISER}
\affiliation{\instIITBhubaneswar}
\affiliation{\instIITGuwahati}
\affiliation{\instIITHyderabad}
\affiliation{\instIITMadras}
\affiliation{\instIndiana}
\affiliation{\instIHEPRussia}
\affiliation{\instHEPHYVienna}
\affiliation{\instIHEPChina}
\affiliation{\instIPP}
\affiliation{\instIOP}
\affiliation{\instIFIC}
\affiliation{\instFrascati}
\affiliation{\instNapoliINFN}
\affiliation{\instPadovaINFN}
\affiliation{\instPerugiaINFN}
\affiliation{\instPisaINFN}
\affiliation{\instRomaINFN}
\affiliation{\instRomaTreINFN}
\affiliation{\instTorinoINFN}
\affiliation{\instTriesteINFN}
\affiliation{\instJAEA}
\affiliation{\instMainz}
\affiliation{\instGiessen}
\affiliation{\instKarlsruhe}
\affiliation{\instKitasato}
\affiliation{\instKISTI}
\affiliation{\instKorea}
\affiliation{\instKSU}
\affiliation{\instKyungpook}
\affiliation{\instLPI}
\affiliation{\instLNNU}
\affiliation{\instLMU}
\affiliation{\instLuther}
\affiliation{\instMNITJaipur}
\affiliation{\instMPP}
\affiliation{\instMPGHLL}
\affiliation{\instMcGill}
\affiliation{\instMEPhI}
\affiliation{\instNagoya}
\affiliation{\instNagoyaKMI}
\affiliation{\instNagoyaIAR}
\affiliation{\instNaraWu}
\affiliation{\instNTUTaiwan}
\affiliation{\instNUUTaiwan}
\affiliation{\instKrakow}
\affiliation{\instNiigata}
\affiliation{\instNSU}
\affiliation{\instOkinawa}
\affiliation{\instOsakaCity}
\affiliation{\instRCNP}
\affiliation{\instPNNL}
\affiliation{\instPanjab}
\affiliation{\instPeking}
\affiliation{\instPanjabPAU}
\affiliation{\instRIKENMSL}
\affiliation{\instSeoul}
\affiliation{\instSPU}
\affiliation{\instSoochow}
\affiliation{\instSoongsil}
\affiliation{\instLjubljanaJSI}
\affiliation{\instKyiv}
\affiliation{\instTata}
\affiliation{\instTUM}
\affiliation{\instTelAviv}
\affiliation{\instToho}
\affiliation{\instTohoku}
\affiliation{\instTitech}
\affiliation{\instTokyoMetropolitan}
\affiliation{\instUAS}
\affiliation{\instNapoliUNIV}
\affiliation{\instPadovaUNIV}
\affiliation{\instPerugiaUNIV}
\affiliation{\instPisaUNIV}
\affiliation{\instRomaUNIV}
\affiliation{\instRomaTreUNIV}
\affiliation{\instTorinoUNIV}
\affiliation{\instTriesteUNIV}
\affiliation{\instMontreal}
\affiliation{\instIJCLab}
\affiliation{\instIPHC}
\affiliation{\instAdelaide}
\affiliation{\instBonn}
\affiliation{\instUBC}
\affiliation{\instCincinnati}
\affiliation{\instFlorida}
\affiliation{\instHawaii}
\affiliation{\instHeidelberg}
\affiliation{\instLjubljanaUniLJ}
\affiliation{\instLouisville}
\affiliation{\instMalaya}
\affiliation{\instLjubljanaUM}
\affiliation{\instMelbourne}
\affiliation{\instMississippi}
\affiliation{\instUOM}
\affiliation{\instPittsburgh}
\affiliation{\instUSTC}
\affiliation{\instSAlabama}
\affiliation{\instSCarolina}
\affiliation{\instSydney}
\affiliation{\instUTokyo}
\affiliation{\instIPMU}
\affiliation{\instVictoria}
\affiliation{\instVPI}
\affiliation{\instWayneState}
\affiliation{\instYamagata}
\affiliation{\instYerevan}
\affiliation{\instYonsei}

  \author{F.~Abudin{\'e}n}\affiliation{\instTriesteINFN} 
  \author{I.~Adachi}\affiliation{\instKEK}\affiliation{\instSOKENDAI} 
  \author{R.~Adak}\affiliation{\instFudan} 
  \author{K.~Adamczyk}\affiliation{\instKrakow} 
  \author{P.~Ahlburg}\affiliation{\instBonn} 
  \author{J.~K.~Ahn}\affiliation{\instKorea} 
  \author{H.~Aihara}\affiliation{\instUTokyo} 
  \author{N.~Akopov}\affiliation{\instYerevan} 
  \author{A.~Aloisio}\affiliation{\instNapoliUNIV}\affiliation{\instNapoliINFN} 
  \author{F.~Ameli}\affiliation{\instRomaINFN} 
% ...
\collaboration{Belle II Collaboration}






\begin{abstract}

We present the results of the re-discovery of the decay $B^0 \to \pi^- \ell^+ \nu_\ell$ in 34.6\invfb of Belle II data using hadronic $B$-tagging via the Full Event Interpretation algorithm. We observe 21 signal events on a background of 155 in a fit to the distribution of the square of the missing mass, $M_{\mathrm{miss}}^2$, with a significance of 5.69$\sigma$, and determine a total branching fraction of (1.58 $\pm$ 0.43$_{\mathrm{stat}}$ $\pm$ 0.07$_{\mathrm{sys}}$) $\times 10^{-4}$.


\end{abstract}


\maketitle

{\renewcommand{\thefootnote}{\fnsymbol{footnote}}}
\setcounter{footnote}{0}


\section{Introduction}

Since the start of its first physics run in 2019, the Belle II detector has collected over 74\invfb of data from electron-positron collisions. This early data has been invaluable for investigating the performance of the detector and the analysis software, and has also led to the re-discovery of many physics processes.

In this paper, we study rare semi-leptonic decays of the form $B^0 \to \pi^- \ell^+ \nu_\ell$, where $\ell = e$, $\mu$ \footnote{Charge conjugate processes are implied for all quoted decays of $B$-mesons throughout this paper.}. These decays are considered golden modes for precise determinations of the absolute value of the CKM-matrix element $|V_{\mathrm{ub}}|$. The integrated luminosity collected at present is too small to provide a competitive measurement. We present the re-discovery of the decay $B^0 \to \pi^- \ell^+ \nu_\ell$ in 34.6\invfb of Belle II data via hadronic $B$-tagging provided by the Full Event Interpretation (FEI) algorithm \cite{Keck:2018lcd}.


\section{The Belle II Detector}
The structure of the Belle II detector is described in detail in Ref. \cite{Abe:2010sj}. The inner-most layers are known collectively as the vertex detector, or VXD, and are dedicated to the tracking of charged particles and the precise determination of particle decay vertices. The VXD is comprised of two layers of silicon pixel sensors surrounded by four layers of silicon strip detectors. The central drift chamber (CDC) surrounds the VXD, encompassing the barrel region of the detector, and is primarily responsible for the reconstruction of charged tracks and the determination of their momenta.

Particle identification is provided by two independent Cherenkov-imaging components, the Time-Of-Propagation (TOP)
counter and the Aerogel Ring-Imaging Cherenkov detector (ARICH), located in the barrel and forward endcap regions of the detector, respectively. The electromagnetic calorimeter (ECL) encases all of the previous layers, and consists of a total of 8736 scintillator crystals used primarily for the determination of the energies of charged tracks and photons. A superconducting solenoid surrounds the inner components and provides the 1.5T magnetic field required by the various sub-detectors. Finally, the $K^0_L$- and muon detector (KLM) forms the outermost detector layer aimed at the detection of $K^0_L$ mesons and muons.

\section{Datasets}
\label{sec:datasets}

The amount of data studied for this analysis corresponds to an integrated luminosity of 34.6\invfb. To simulate signal and background, fully simulated Monte Carlo (MC) samples of charged and neutral decays of pairs of $B$-mesons, as well as 
continuum \epem \to \qqbar ($q = u,\,d,\,s,\,c$) were used, with all samples generated alongside beam background effects including beam scattering and radiative processes. Table \ref{table:MC13} lists the number of events used for each of the MC components. These samples correspond to a total integrated luminosity of 200\invfb. 

\begin{table} [h!]
\def\arraystretch{1.2}%
\begin{tabular}{|c|c|}
\hline
&   $N_{\mathrm{events}}$ used ($\times 10^6$) \\\hline
 $B^+B^-$ &  108.0 \\\hline
 $B^0\bar{B}^0$ &  102.0 \\\hline
 $c\bar{c}$ &  265.8 \\\hline
 $u\bar{u}$ &  321.0 \\\hline
 $s\bar{s}$ &  76.6 \\\hline
 $d\bar{d}$ &  80.2 \\\hline
\end{tabular}
\caption{Number of MC events used for the analysis, equivalent to 200\invfb.}\label{table:MC13}
\end{table}

In addition to the generic MC, dedicated $B \to \X_u \ell \nu$ signal samples were used, where $X_u$ is a hadronic system resulting from the quark flavor transition $b \to u$, for the in-depth study of the signal decays and related backgrounds, and included the implementation of the hybrid modelling technique for $b \to u$ transitions described in Ref. \cite{Ramirez:1990db}. Each $B^+ \to \X_u \ell \nu$ and $B^0 \to \X_u \ell \nu$ sample consisted of a total of 50 million resonant (R) events containing the relevant exclusive decays as well as 50 million non-resonant (I) events corresponding to the inclusive component, simulated using the BLNP heavy-quark-effective-theory-based model \cite{Lange:2005ll}. These samples were then combined together and the eFFORT tool \cite{markus_prim_2020_3965699} was used to calculate a hybrid weight per event in three-dimensional bins of the generated lepton energy in the $B_{\mathrm{sig}}$-frame, $E^B_\ell$, the four-momentum transfer to the leptonic system, $q^2$, and the mass of the hadronic system containing an up-quark, $M_X$, such that $H_i = R_i + w_i I_i.$ The number of total hybrid events per bin, $H_i$, is the sum of the number of resonant events $R_i$ and the number of inclusive events $I_i$ scaled down by the appropriate weight $w_i$. 

For the analysis, the $B \to \X_u \ell \nu$ events from the generic 200\invfb MC sample were removed and replaced with the equivalent amount of this hybrid re-weighted MC.

\section{Full Event Interpretation}
\label{sec:FEISkim}

\begin{figure}[h!]
\begin{center}
\includegraphics[scale=0.8]{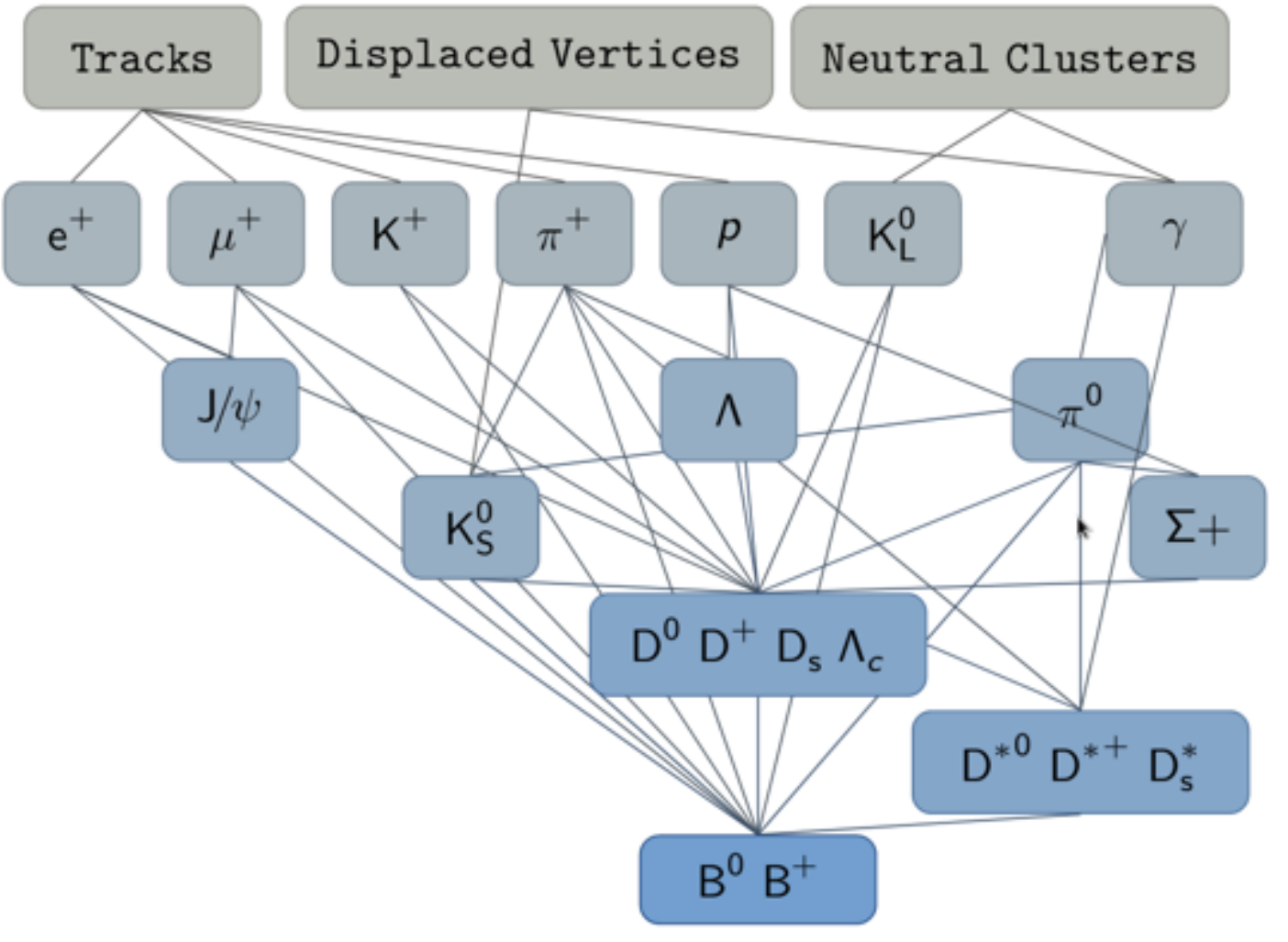}
  \caption{The hierarchical structure of the Full Event Interpretation tagging algorithm. Detector information
including tracks and energy deposits in the electromagnetic calorimeter (ECL) and $K^0_L$ and muon detector (KLM) is first used to reconstruct and train a classifier for each final-state particle. Intermediate particles are then reconstructed in stages to form the parent $B$-mesons, each with an associated \texttt{SignalProbability}.}
  \label{fig:fei}
\end{center}
\end{figure}

The Full Event Interpretation (FEI) \cite{Keck:2018lcd} is a machine learning algorithm developed for tagged analysis at Belle II. It supports both hadronic and semi-leptonic tagging, reconstructing $B$-mesons across more than 4000 individual decay chains. The algorithm utilises the FastBDT software package that trains a series of multi-variate classifiers for each tagging channel via a number of stochastic gradient-boosted decision trees \cite{Keck:2016tk}. The training is performed in a hierarchical manner with final-state particles being reconstructed first from detector information. The decay channels are then built up from these particles as illustrated in Figure \ref{fig:fei}, with the reconstruction of the $B$-mesons performed last. For each $B$-meson tag candidate reconstructed by the FEI, a value of the final multi-variate classifier output, the \texttt{SignalProbability}, is assigned. The \texttt{SignalProbability} is distributed between 0 and 1, representing candidates identified as being background-like and signal-like, respectively.\\
Official FEI skims of both data and MC are produced centrally by the Belle II collaboration, and are available for use in analyses. These include both hadronic and semileptonic skims, and involve the application of the FEI together with a number of loose selections that aim to reduce the sample sizes with little to no loss of signal events. 

For the hadronic FEI, the minimum number of charged particle tracks per event satisfying certain quality criteria is set to three. The vast majority of $B$-meson decay chains corresponding to the hadronic FEI channels include at least three charged particles, and such a cut is useful at suppressing background from non-$B\bar{B}$ events. The criteria chosen include track parameter cuts to ensure close proximity to the interaction point (IP), with the distance from the center of the detector along the $z$-axis (corresponding to the solenoidal-field axis) and in the transverse plane satisfying $|z_0| < 2.0\,\rm cm$ and $|d_0| < 0.5\,\rm cm$, respectively. A minimum threshold $p_t >$ 0.1 GeV is placed on the track transverse momentum \footnote{Throughout this paper, we use the convention $c=1$ for the speed of light.}. Similar restrictions are applied to the ECL clusters in the event, with at least three clusters required within the polar angle acceptance of the CDC, 0.297 $< \theta <$ 2.618, that satisfy a minimum energy threshold E $>$ 0.1 GeV. The total visible energy per event is required to be at least 4 GeV. The total energy deposited in the ECL is restricted to $2\,{\rm GeV} < E_{\rm ECL} < 7\,{\rm GeV}$, however, to suppress events with an excess of energy deposits due to beam background.

The FEI typically results in many $B_{\mathrm{tag}}$ candidates per event. The number of these candidates is reduced with selections on the beam-constrained mass, $M_{\mathrm{bc}}$, and energy difference, $|\Delta E|$, with these defined as follows:\\

$$M_{\mathrm{bc}} = \sqrt{\frac{E_{\mathrm{beam}}^{2}}{4}-\vec{p}^{\,2}_{B_{\mathrm{tag}}}}\hspace{0.5em},  \hspace{3em} \Delta E = E_{B_{\mathrm{tag}}} - \frac{E_{\mathrm{beam}}}{2},$$
where $E_{\mathrm{beam}}$ is the centre-of-mass (CMS) energy of the $e^+e^-$ system, 10.58 GeV, and $\vec{p}_{B_{\mathrm{tag}}}$ and $E_{B_{\mathrm{tag}}}$ are the  $B_{\mathrm{tag}}$ momentum and energy in the CMS frame, respectively. The cuts applied during the hadronic FEI skim are $M_{\mathrm{bc}} > 5.24$ GeV and $|\Delta E| < 0.2$ GeV.

Finally, a loose cut on the $B_{\mathrm{tag}}$ classifier output,  \texttt{SignalProbability} $> 0.001$, provides further background rejection with little signal loss. 

\section{Event Selection and Analysis Strategy}
\label{sec:selections}

For this analysis, the distribution of the square of the missing mass, $M_{\mathrm{miss}}^2$, was the variable chosen for the determination of the $B^0 \to \pi^- \ell^+ \nu_\ell$ yield. We define the four-momentum of the signal $B$-meson in the CMS frame as follows,\\

\begin{center}
    $$p_{B_{\mathrm{sig}}} \equiv  (E_{B_{\mathrm{sig}}}, \vec{p}_{B_{\mathrm{sig}}}) = \left(\frac{m_{\Upsilon(4S)}}{2}, -\vec{p}_{B_{\mathrm{tag}}}\right) \hspace{0.5em},$$
    
\end{center}
where $m_{\Upsilon(4S)}$ is the mass of the $\Upsilon$(4S) meson as recorded by the PDG \cite{Tanabashi:2018oca}. We set the energy of $B_{\mathrm{sig}}$ to be half of the $\Upsilon$(4S) rest mass, and take the $B_{\mathrm{sig}}$ momentum to be the negative $B_{\mathrm{tag}}$ momentum. We then define the missing four-momentum as

\begin{center}
    $p_{\mathrm{miss}} \equiv (E_{\mathrm{miss}}, \vec{p}_{\mathrm{miss}}) =  p_{B_{\mathrm{sig}}} - p_Y$,
\end{center}
where $Y$ is a pseudo-particle formed from the combination of the pion and lepton four-momenta. The square of the missing momentum can then simply be defined as $M_{\mathrm{miss}}^2 \equiv p_{\mathrm{miss}}^2$.

The signal selection criteria were determined using simulation, with data blinded in the signal region of $M_{\mathrm{miss}}^2 \leq 1.0$ GeV$^2$ and the agreement between data and MC was investigated in the $M_{\mathrm{miss}}^2$ sideband. The event selections applied follow closely those from a 2013 study \cite{Sibidanov:2013sb} of exclusive $B \to \X_u \ell \nu$ decays using the full 711\invfb Belle dataset with hadronic tagging. All selections were applied in addition to the hadronic FEI skim cuts detailed in the previous section.

At the event-level, a loose selection on the second normalised Fox-Wolfram moment \cite{Wolfram:1978fw} was applied to suppress continuum background, with R2 $<$ 0.4. To reject incorrectly reconstructed $B_{\mathrm{tag}}$ candidates, the tag-side beam-constrained mass cut was tightened to $M_{\mathrm{bc}} > 5.27$ GeV. The $B_{\mathrm{tag}}$ candidate having the highest value of the \texttt{SignalProbability} classifier output was retained in each event.

For the reconstructed electrons and muons, cuts on the track impact parameters were applied to select those originating close to the interaction point (IP). These consisted of cuts on the $z$-axis and transverse-plane distances from the IP of $|dz| < 5\,{\rm cm}$ and $dr < 2$ cm, respectively. Only those leptons within the acceptance of the CDC were selected. Electrons and muons were identified through cuts on the particle identification variables provided by the Belle II Analysis Software Framework (BASF2) \cite{Kuhr:B2}. These variables describe the likelihood that a charged track corresponds to a specified final-state particle against all other possible particles, and are defined between 0 (unlikely) and 1 (likely). The likelihoods are built from a combination of the information returned from all of the individual sub-detectors. The relevant variables and their selections were \texttt{electronID} $> 0.9$ and \texttt{muonID} $> 0.9$. A minimum threshold on the lab-frame momentum was placed on the reconstructed leptons, with $p_{\mathrm{lab}} > 0.3$ GeV for electrons and $p_{\mathrm{lab}} > 0.6$ GeV for muons. The four-momentum of the reconstructed electrons was also corrected in order to account for Bremsstrahlung radiation. For any energy deposit in the ECL not associated with a charged particle track, it was checked whether the distance between this cluster and the cluster associated with an electron track was less than 0.5 mm. If this condition was met and the cluster energy was between 0.2 and 1 times the energy of the track assigned to the electron, a weight was calculated for the relation between the electron and photon. In the case of multiple photons satisfying these conditions, only the photon with the weight corresponding to the closest relation was taken, its four-momentum was added to the electron, and the photon was excluded from the rest of the event. Finally, a single lepton was kept per event with the highest value of the \texttt{electronID} or \texttt{muonID}.

For the reconstructed charged pions, similar impact parameter cuts were applied as those for the leptons, with $dr < 2$ cm and $|dz| < 4$ cm. The charged pion tracks were similarly only selected within the CDC acceptance, with the additional condition that a minimum number of 20 hits was recorded by the CDC. The sign of the charge of the reconstructed pion was explicitly required to be opposite that of the lepton. A cut on the relevant particle identification variable was also applied, with \texttt{pionID} $> 0.5$.

Combining the centre-of-mass frame four-momenta of the reconstructed pion and lepton  together into the pseudo-particle $Y$, a selection on the cosine of the angle between the flight direction of the signal $B$-meson and the $Y$, $cos(\theta_{BY})$, was made. A value of $|cos(\theta_{BY})| < 1$ is expected if only a neutrino is missing in the reconstruction. However, to account for resolution effects and to avoid introducing potential bias in the background $M_{\mathrm{miss}}^2$ distributions, this requirement was loosened to $|cos(\theta_{BY})| < 3$. To ensure that the reconstructed lepton and pion tracks originated from the same vertex, a cut on the difference between the $z$-coordinates of both tracks was applied, with $|z_\ell - z_\pi| < 1$ mm.

For this analysis, a single $\Upsilon$(4S) candidate with the lowest value of the $M_{\mathrm{miss}}^2$ was retained per event. All remaining tracks and clusters in the ECL after the reconstruction of the $\Upsilon$(4S) were combined into a single object known as the rest-of-event (ROE). Events in which additional tracks satisfying the conditions $dr < 2$ cm, $|dz| < 5$ cm and $p_t > 0.2$ GeV remained after the reconstruction of the $\Upsilon$(4S) were excluded. The CMS energies in the ROE corresponding to deposits of neutral particles in the ECL were summed for those clusters satisfying energy cuts of $E > 0.1$ GeV, $E > 0.09$ GeV and $E > 0.16$ GeV for the forward end-cap, barrel and backward end-cap regions respectively. A minimum threshold on this energy was placed to account for the neutrino, with $E_{\mathrm{miss}} > 0.3$ GeV. The extra energy was also required to be below a maximum value of $E_{\mathrm{residual}} < 1.0$ GeV.




\section{Results}
\label{sec:results}


Before unblinding the dataset, a toy MC study was performed to evaluate the expected significance of the $B^0 \to \pi^-\ell^+\nu_\ell$ signal in 34.6\invfb. To evaluate the number of events expected in data, the number of total MC events normalised to this integrated luminosity and passing all selections was scaled down by a hadronic FEI calibration factor of 0.8301. This factor was applied in order to account for the difference in the tag-side reconstruction efficiency of the FEI between data and MC. An independent study was performed in order to evaluate this factor through fitting the lepton momentum spectrum in $B\to X \ell\nu$ decays and taking the ratio of signal events in data and MC.

Template Probability Density Functions (PDFs) were built from signal and background $M_{\mathrm{miss}}^2$ distributions in MC, with signal candidates defined as those reconstructed from MC events containing $B^0 \to \pi^- \ell^+ \nu_\ell$ decays. The various backgrounds included the cross-feeds from $B^0 \to \rho^- \ell^+ \nu_\ell$ and other $B \to X_u \ell \nu$ decays, as well as candidates reconstructed from other generic $B\bar{B}$ and continuum events. Due to low statistics, these background components were combined together into a single background PDF. Each MC template was weighted by a set of corrections to account for the difference in lepton identification efficiencies and pion and kaon fake-rates between MC and data. These corrections were evaluated per event based on the magnitude of the lab-frame momentum $p$ and polar angle $\theta$ of the reconstructed lepton tracks.

\begin{figure}[h!]
\begin{center}
\includegraphics[scale=0.35]{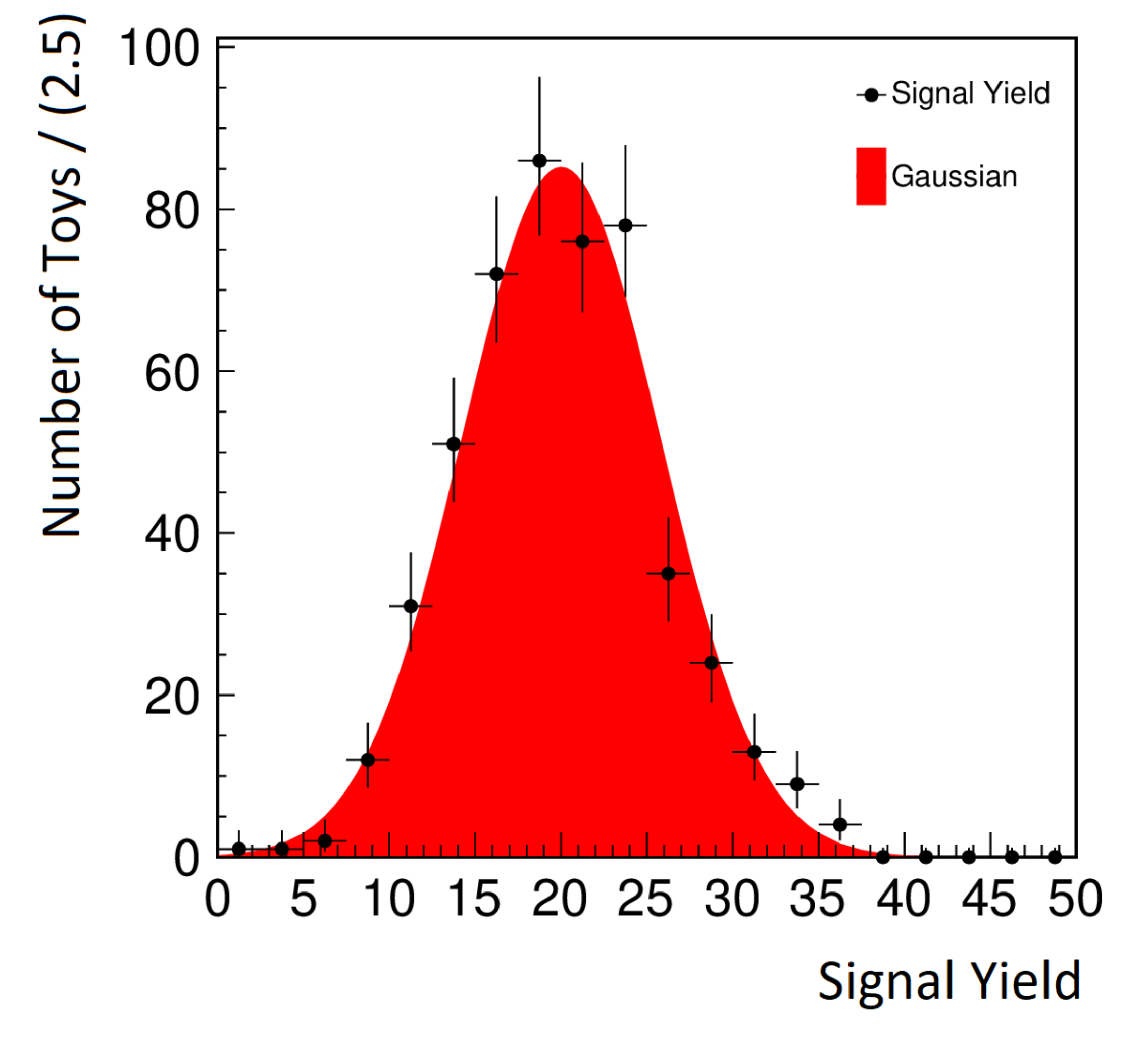}
  \caption{Distribution of the fitted signal yields derived from fits to the $M_{\mathrm{miss}}^2$ distributions of 500 toy MC $B^0 \to \pi^-\ell^+\nu_\ell$ samples.} 
  \label{fig:Btoy}
\end{center}
\end{figure}

A set of 500 toy MC samples was generated from the combined signal and background template distributions. Each toy consisted of a $M_{\mathrm{miss}}^2$ distribution with a total number of events defined  within a range of Poissonian uncertainty on the expected number of data events. Under the signal + background hypothesis, extended unbinned maximum likelihood template fits to the $M_{\mathrm{miss}}^2$ distributions of each toy were performed using the RooFit framework  \cite{Verkerke:2003ir}, with all background components combined into a single background template. A Gaussian PDF was fitted to the signal yields obtained from the fits to each toy sample and is shown in Figure \ref{fig:Btoy}. The mean and standard deviation of the fitted Gaussian is listed in Table \ref{table:significance}, together with the expected yield as determined from the MC. The mean signal yield agrees with the predicted yield within uncertainty.

Additional extended maximum likelihood fits were then performed to the same toy MC samples under the background-only hypothesis. The likelihood ratio $\lambda$ between both fits was computed for each toy:

$$\lambda = \frac{\mathcal{L}_{B}}{\mathcal{L}_{S+B}} \hspace{0.5em},$$
where $\mathcal{L}_{B}$ and $\mathcal{L}_{S+B}$ are the maximised likelihoods returned by the fits to the background-only and signal + background hypotheses, respectively.

A significance estimator $\Sigma$ was subsequently defined for each toy sample based on the likelihood ratio:

$$\Sigma = \sqrt{-2\ln\lambda}\hspace{0.5em} .$$

\begin{figure}[h!]
\begin{center}
\includegraphics[scale=0.4]{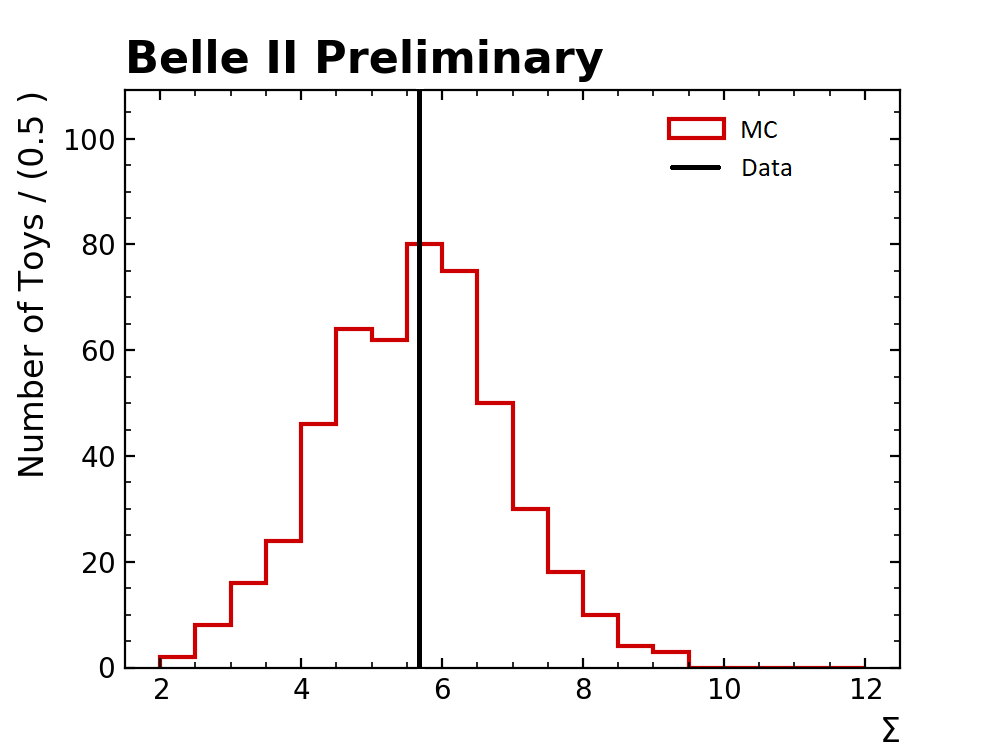}
  \caption{Distribution of the estimator of expected significance for $B^0 \to \pi^- \ell^+ \nu_\ell$ decays across 500 toy MC samples. The observed significance in data is also plotted.} 
  \label{fig:teststat}
\end{center}
\end{figure}

Figure \ref{fig:teststat} shows the resultant distribution of $\Sigma$ for the decays to charged pions. The median of the distribution corresponds to a significance of 5.63$\sigma$.

\begin{figure}[h!]
\begin{center}
\includegraphics[scale=0.9]{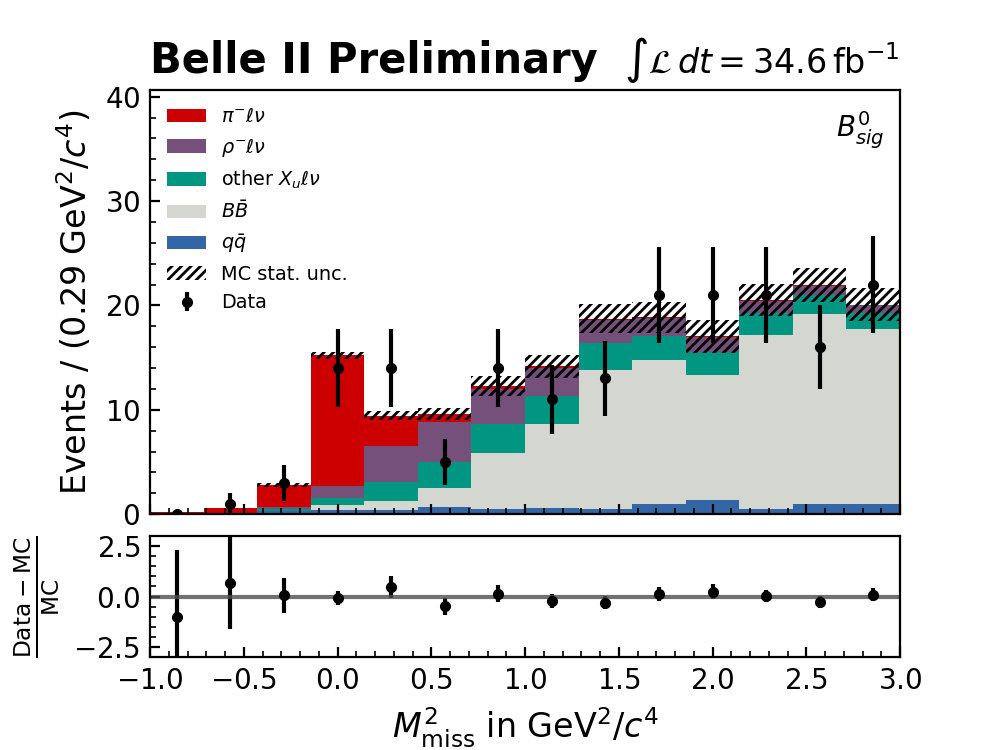}
  \caption{Pre-fit $M_{\mathrm{miss}}^2$ distribution from 34.6\invfb of data. The MC has been scaled down by a hadronic FEI calibration factor of 0.8301.}
  \label{fig:B0SP001unblind}
\end{center}
\end{figure}

The data was subsequently unblinded in the signal region of $M_{\mathrm{miss}}^2 \leq 1.0$ GeV$^2$, with the resultant unfitted distribution displayed in Figure \ref{fig:B0SP001unblind}. For charged particle tracks in data, a factor of 1.00056 was applied to the track momentum to correct for differences between data and MC. In addition to the lepton ID corrections, each MC component was scaled down by the hadronic FEI calibration factor of 0.8301 in order to account for the difference in the FEI reconstruction efficiency between data and MC. Fairly good agreement was observed across the $M_{\mathrm{miss}}^2$ distribution, including in the signal region. A clear signal peak can be seen at $M_{\mathrm{miss}}^2$ = 0 for both data and MC, with all other backgrounds peaking at higher values of $M_{\mathrm{miss}}^2$.

\begin{figure}[h!]
\begin{center}
\includegraphics[scale=0.9]{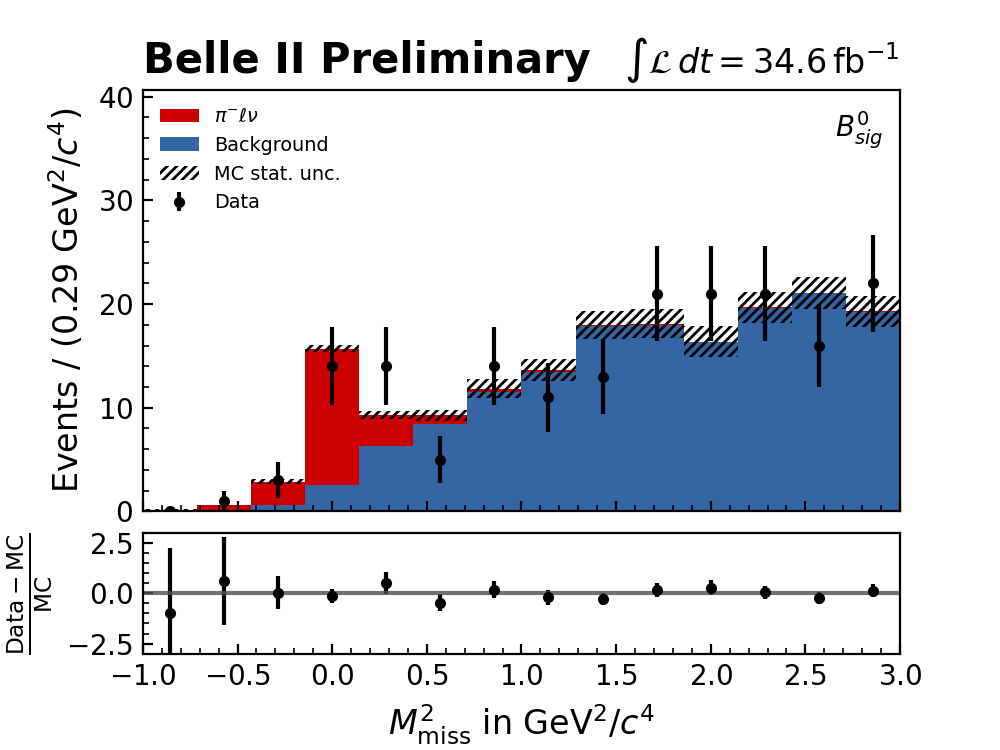}
  \caption{Post-fit $M_{\mathrm{miss}}^2$ distribution from 34.6\invfb of data.}
  \label{fig:B0SP001fit}
\end{center}
\end{figure}

The signal and background MC templates shown in Figure \ref{fig:B0SP001unblind} were then used to perform an extended unbinned maximum likelihood fit to the $M_{\mathrm{miss}}^2$ distribution in data under the signal + background hypothesis. Figure \ref{fig:B0SP001fit} displays this fitted distribution and the signal yield obtained is listed in Table \ref{table:significance}. The observed yield in data agrees well with the predicted yield from MC.

The observed significance was similarly evaluated by performing an additional fit to data under the background-only hypothesis. A significance of $5.69\sigma$ was observed, as shown in Figure \ref{fig:teststat}.

Finally, the branching fraction for the $B^0 \to \pi^- \ell^+ \nu_\ell$ decay was extracted using the following formula:\\

$$\mathcal{B}(B^0\to\pi^-\ell^+\nu_\ell) = \frac{N_{\mathrm{sig}}^{\mathrm{data}}(1 + f_{\mathrm{+0}})}{4\times \mathrm{CF}_{\mathrm{FEI}} \times N_{B\bar{B}} \times \epsilon} \hspace{0.5em},$$
where $N_{\mathrm{sig}}^{\mathrm{data}}$ is the fitted signal yield obtained from data, $f_{\mathrm{+0}}$ is the ratio between the branching fractions of the decays of the $\Upsilon$(4S) meson to pairs of charged and neutral $B$-mesons \cite{Tanabashi:2018oca}, $\mathrm{CF}_{\mathrm{FEI}}$ is the FEI calibration factor, $N_{B\bar{B}}$ is the number of $B$-meson pairs counted in the current dataset, and $\epsilon$ is the reconstruction efficiency. The factor of 4 present in the denominator accounts for the two $B$-mesons in the $\Upsilon$(4S) decay and the reconstruction of both light lepton flavors.

The signal efficiency was calculated from the ratio of signal events present in MC before and after all analysis selections, and was determined to be (0.216 $\pm$ 0.001)$\%$. The values of the above parameters together with the measured branching fraction are summarised in Table \ref{table:BF}. The branching fraction agrees well with the world average value, (1.50 $\pm$ 0.06) $\times 10^{-4}$ \cite{Tanabashi:2018oca}.

\begin{table} [h!]
\def\arraystretch{1.1}%

 
\begin{tabular}{|p{7em}|p{7em}|p{7em}|p{7em}|}
\hline
Predicted yield from MC & Fitted mean $\mu$ from MC & Fitted standard deviation $\sigma$ from MC & Observed fitted yield in data\\\hline
19.83 & 20.01 $\pm$ 0.26 & 5.79 $\pm$ 0.19 & 20.79 $\pm$ 5.68\\\hline

\end{tabular}
\caption{Mean and standard deviation of the Gaussian PDF fit to the distributions of signal yields from 500 toy MC samples. The numbers of expected and observed signal events in data are also listed. }\label{table:significance}
\end{table}

\begin{table} [h!]
\def\arraystretch{1.2}%
\begin{tabular}{|c|c|}
\hline
 $N_{\mathrm{sig}}^{\mathrm{data}}$ & 20.79 $\pm$ 5.68  \\\hline
 $f_{\mathrm{+0}}$ &  1.058 $\pm$ 0.024 \\\hline
 $\mathrm{CF}_{\mathrm{FEI}}$ &  0.8301 $\pm$ 0.0286 \\\hline
 $N_{B\bar{B}}$ & (37.711 $\pm$ 0.602) $\times 10^6$  \\\hline
 $\epsilon$ &  (0.216 $\pm$ 0.001)$\%$ \\\hline
 $\mathcal{B}(B^0\to\pi^-\ell^+\nu_\ell)$ & (1.58 $\pm$ 0.43$_{\mathrm{stat}}$ $\pm$ 0.07$_{\mathrm{sys}}$) $\times 10^{-4}$\\\hline
\end{tabular}
\caption{Measured branching fraction of $B^0 \to \pi^- \ell^+ \nu_\ell$ decays using 34.6\invfb of data. The values of the parameters used in the measurement are also given.}\label{table:BF}
\end{table}



\section{Systematic Uncertainties}
\label{sec:systematics}

A number of sources of systematic uncertainty were identified for this analysis and evaluated for the branching fraction measurement. The relative uncertainties for each source, in percent, are summarised in Table \ref{table:syserrors}, and included:

\begin{itemize}
    \item \boldmath $f_{\mathrm{+0}}$: \unboldmath We combine the errors on the world averages for the branching fractions $\mathcal{B}(\Upsilon$(4S)$\to B^+B^-$) and $\mathcal{B}(\Upsilon$(4S) $\to B^0\bar{B}^0$) and calculate the relative uncertainty on the fraction $f_{\mathrm{+0}}$.
    \item \textbf{FEI calibration:} The given uncertainty on the calibration factor for the hadronic FEI was determined taking into account multiple systematic effects in the fitting to the lepton momentum spectrum of $B\to X \ell \nu$ decays. These included uncertainties on both the branching fractions and form factors of the various semileptonic components of $B\to X \ell \nu$, the lepton ID efficiency and fake rate uncertainties, tracking uncertainties, and template uncertainties as a result of MC statistics. The relative uncertainty on the calibration factor forms the dominant source of systematic uncertainty for this analysis.
    \item \boldmath $N_{B\bar{B}}$: \unboldmath The uncertainty on the number of $B\bar{B}$ events in the present dataset includes systematic effects due to uncertainties on the luminosity, beam energy spread and shift, tracking efficiency and the selection efficiency of $B\bar{B}$ events.
    \item \textbf{Reconstruction efficiency:} We represent the uncertainty on the reconstruction efficiency with a binomial error dependent on the size of the MC samples used for the analysis.  
    \item \textbf{Tracking:} We assign a flat systematic uncertainty of 0.80$\%$ for each charged track. We assume complete correlation between the uncertainties of the lepton and pion tracks and obtain a total relative uncertainty of 1.60$\%$.
    \item \textbf{Lepton identification:} The lepton efficiencies and pion fake rates are evaluated in bins of the lepton momentum $p$ and polar angle $\theta$, each with statistical and systematic uncertainties. The effect of these uncertainties on the signal reconstruction efficiency was determined through generating 500 variations on the nominal correction weights via Gaussian smearing. The relative uncertainty was then taken from the spread on the values of the reconstruction efficiency over all variations. 
\end{itemize}
\begin{table} [h!]
\def\arraystretch{1.2}%
\begin{tabular}{|c|c|}
\hline
Source of systematic uncertainty &   $\%$ of $\mathcal{B}$ \\\hline\hline
 $f_{\mathrm{+0}}$ & 1.17 \\\hline
 FEI calibration &  3.45 \\\hline
 $N_{B\bar{B}}$ &  1.60 \\\hline
 Reconstruction efficiency $\epsilon$ &  0.46 \\\hline
 Tracking & 1.60 \\\hline
 Lepton ID & 1.05\\\hline\hline
 Total & 4.44\\\hline
\end{tabular}
\caption{Sources of systematic uncertainties and their percentages of the total measured branching fraction.}\label{table:syserrors}
\end{table}

We do not currently include systematic uncertainties due to modelling $B \to X_u \ell \nu$ decays, as these are expected to be sufficiently small in comparison with the other sources listed for the current dataset.

\section{Summary}
\label{sec:conclusions}
In summary, we have presented a re-discovery of the semi-leptonic decay $B^0 \to \pi^- \ell^+ \nu_\ell$ via hadronic tagging in 34.6\invfb of Belle II data, with an observed significance of 5.69$\sigma$. A branching fraction of (1.58 $\pm$ 0.43$_{\mathrm{stat}}$ $\pm$ 0.07$_{\mathrm{sys}}$) $\times 10^{-4}$ was measured for this decay, in agreement with the world average \cite{Tanabashi:2018oca}.

\section*{Acknowledgements}
\label{sec:acknowledgements}
We thank the SuperKEKB group for the excellent operation of the
accelerator; the KEK cryogenics group for the efficient
operation of the solenoid; and the KEK computer group for
on-site computing support.
This work was supported by the following funding sources:
Science Committee of the Republic of Armenia Grant No. 18T-1C180;
Australian Research Council and research grant Nos.
DP180102629, 
DP170102389, 
DP170102204, 
DP150103061, 
FT130100303, 
and
FT130100018; 
Austrian Federal Ministry of Education, Science and Research, and
Austrian Science Fund No. P 31361-N36; 
Natural Sciences and Engineering Research Council of Canada, Compute Canada and CANARIE;
Chinese Academy of Sciences and research grant No. QYZDJ-SSW-SLH011,
National Natural Science Foundation of China and research grant Nos.
11521505,
11575017,
11675166,
11761141009,
11705209,
and
11975076,
LiaoNing Revitalization Talents Program under contract No. XLYC1807135,
Shanghai Municipal Science and Technology Committee under contract No. 19ZR1403000,
Shanghai Pujiang Program under Grant No. 18PJ1401000,
and the CAS Center for Excellence in Particle Physics (CCEPP);
the Ministry of Education, Youth and Sports of the Czech Republic under Contract No.~LTT17020 and 
Charles University grants SVV 260448 and GAUK 404316;
European Research Council, 7th Framework PIEF-GA-2013-622527, 
Horizon 2020 Marie Sklodowska-Curie grant agreement No. 700525 `NIOBE,' 
and
Horizon 2020 Marie Sklodowska-Curie RISE project JENNIFER2 grant agreement No. 822070 (European grants);
L'Institut National de Physique Nucl\'{e}aire et de Physique des Particules (IN2P3) du CNRS (France);
BMBF, DFG, HGF, MPG, AvH Foundation, and Deutsche Forschungsgemeinschaft (DFG) under Germany's Excellence Strategy -- EXC2121 ``Quantum Universe''' -- 390833306 (Germany);
Department of Atomic Energy and Department of Science and Technology (India);
Israel Science Foundation grant No. 2476/17
and
United States-Israel Binational Science Foundation grant No. 2016113;
Istituto Nazionale di Fisica Nucleare and the research grants BELLE2;
Japan Society for the Promotion of Science,  Grant-in-Aid for Scientific Research grant Nos.
16H03968, 
16H03993, 
16H06492,
16K05323, 
17H01133, 
17H05405, 
18K03621, 
18H03710, 
18H05226,
19H00682, 
26220706,
and
26400255,
the National Institute of Informatics, and Science Information NETwork 5 (SINET5), 
and
the Ministry of Education, Culture, Sports, Science, and Technology (MEXT) of Japan;  
National Research Foundation (NRF) of Korea Grant Nos.
2016R1\-D1A1B\-01010135,
2016R1\-D1A1B\-02012900,
2018R1\-A2B\-3003643,
2018R1\-A6A1A\-06024970,
2018R1\-D1A1B\-07047294,
2019K1\-A3A7A\-09033840,
and
2019R1\-I1A3A\-01058933,
Radiation Science Research Institute,
Foreign Large-size Research Facility Application Supporting project,
the Global Science Experimental Data Hub Center of the Korea Institute of Science and Technology Information
and
KREONET/GLORIAD;
Universiti Malaya RU grant, Akademi Sains Malaysia and Ministry of Education Malaysia;
Frontiers of Science Program contracts
FOINS-296,
CB-221329,
CB-236394,
CB-254409,
and
CB-180023, and SEP-CINVESTAV research grant 237 (Mexico);
the Polish Ministry of Science and Higher Education and the National Science Center;
the Ministry of Science and Higher Education of the Russian Federation,
Agreement 14.W03.31.0026;
University of Tabuk research grants
S-1440-0321, S-0256-1438, and S-0280-1439 (Saudi Arabia);
Slovenian Research Agency and research grant Nos.
J1-9124
and
P1-0135; 
Agencia Estatal de Investigacion, Spain grant Nos.
FPA2014-55613-P
and
FPA2017-84445-P,
and
CIDEGENT/2018/020 of Generalitat Valenciana;
Ministry of Science and Technology and research grant Nos.
MOST106-2112-M-002-005-MY3
and
MOST107-2119-M-002-035-MY3, 
and the Ministry of Education (Taiwan);
Thailand Center of Excellence in Physics;
TUBITAK ULAKBIM (Turkey);
Ministry of Education and Science of Ukraine;
the US National Science Foundation and research grant Nos.
PHY-1807007 
and
PHY-1913789, 
and the US Department of Energy and research grant Nos.
DE-AC06-76RLO1830, 
DE-SC0007983, 
DE-SC0009824, 
DE-SC0009973, 
DE-SC0010073, 
DE-SC0010118, 
DE-SC0010504, 
DE-SC0011784, 
DE-SC0012704; 
and
the National Foundation for Science and Technology Development (NAFOSTED) 
of Vietnam under contract No 103.99-2018.45.

\bibliography{belle2}
\bibliographystyle{belle2-note}

\end{document}